\documentclass[journal, 10pt]{IEEEtran}

\usepackage{array}
\usepackage{amsmath}
\usepackage{amssymb}
\usepackage{graphicx}
\usepackage{subcaption}[position=b, textformat=period]
\usepackage{makecell}
\usepackage{enumitem}
\usepackage[binary-units=true]{siunitx}
\usepackage{optidef}
\usepackage{hyperref}
\usepackage[table]{xcolor}
\usepackage{multirow}
\usepackage[ruled]{algorithm2e}
\newcommand\notype[1]{\unskip}
\graphicspath{{Figures/}}
\newcommand{\pubkey}{k^{\text{pub}}_\varepsilon}
\newcommand{\prikey}{k^{\text{pri}}_\varepsilon}
\newcommand{\networkc}{\mathcal{N}}
\newcommand{\applicationc}{\mathcal{A}}
\usepackage{mathtools}

\DeclarePairedDelimiter{\ceil}{\lceil}{\rceil}

\begin{document}
	%\title{A Blockchain-based LoRa System with Edge Computing for Data Security of Internet of things}
	\title{Design and Prototype Implementation of a Blockchain-enabled LoRa System with Edge Computing}
	\author{Lu Hou, Kan Zheng~\IEEEmembership{Senior~Member,~IEEE,} Zhiming Liu, Xiaojun Xu, Tao Wu
	\thanks{This work was supported in part by National Natural Science Foundation of China (NSFC) under Grant 61731004, in part by Huawei Technologies CO.,ltd., and in part by the Key Lab of Universal Wireless Communications, Ministry of Education.}
	\thanks{L.~Hou, K.~Zheng, Z.~Liu, X.~Xu and T.~Wu are with the Intelligent Computing and Communications (IC$^2$) Lab, Wireless Signal Processing and Networks Lab (WSPN), Key Lab of Universal Wireless Communications, Ministry of Education, Beijing University of Posts \& Telecommunications, Beijing, China, 100088.}
	\thanks{Corresponding author: Kan~Zheng.}
	\thanks{Contact email: zkan@bupt.edu.cn}
	\thanks{Copyright (c) 20xx IEEE. Personal use of this material is permitted. However, permission to use this material for any other purposes must be obtained from the IEEE by sending a request to pubs-permissions@ieee.org.}}
	\maketitle
	
	\begin{abstract}
	Efficiency and security have become critical issues during the development of Long Range (LoRa) system for Internet of Things (IoT) applications. The centralized work method in LoRa system, where all packages are processed and kept in central cloud, cannot well exploit the resources in LoRa gateways and also makes it vulnerable to security risks such as data falsification or data loss. On the other hand, the blockchain has the potential to provide a decentralized and secure infrastructure for LoRa system. However, there are significant challenges on deploying blockchain at LoRa gateways with limited edge computing abilities. This paper proposes a design and implementation of blockchain-enabled LoRa system with edge computing by using the open-source Hyperledge Fabric, which is called as HyperLoRa. According to different features of LoRa data, a blockchain network with multiple ledgers is designed, each of which stores a specific kind of LoRa data. LoRa gateways can participate in the operations of the blockchain and share the ledgers that keep the time critical network data with small size. Then, the edge computing abilities of LoRa gateways are utilized to handle the join procedure and application packages processing. Furthermore, a HyperLoRa prototype is implemented on embedded hardware, which demonstrates the feasibility of deploying the blockchain into LoRa gateways with limited computing and storage resources. Finally, various experiments are conducted to evaluate the performances of the proposed LoRa system.
	\end{abstract}
	
	
	\section{Introduction}
	Long Range (LoRa), one of the low-power wide-area (LPWA) technologies, is expected to provide energy-efficient communications to a massive number of end-devices that are distributed in a wide area~\cite{LPWA}. It is an attractive solution for various Internet of Things (IoT) applications, such as environment monitoring and Industrial IoT (IIoT)~\cite{rtlora}. In these applications, a great amount of IoT data can be collected by LoRa gateways in a low sampling rate and processed by the central cloud~\cite{IoTcloud}. However, there are a few challenges that have to be solved before the LoRa systems are widely deployed. Most of them are about how to improve the efficiency and security of LoRa systems.
	\par
	The current LoRa gateways are only responsible for the transparent forwarding of packages between end-devices and central cloud. Then, all LoRa packages are handled and stored in the central cloud, which undertakes all of the workloads~\cite{openlora}. Therefore, the computing and storage resources in the LoRa gateways are not well exploited. Moreover, the central cloud is far from the end-devices, making it hard to meet the low-latency requirements of time-critical IoT applications~\cite{iotedge}. The LoRa systems also have their potential risks of security. For example, the LoRa data are vulnerable to falsification and destruction, since all of the data are gathered into central cloud~\cite{iotdevicethreat, IoTattack}. 

	\par
	Thanks to the development of edge computing technologies, some of the workloads of central cloud can be transferred to LoRa gateways, such as the management of end-devices and the basic processing of LoRa packages. Meanwhile, the blockchain with decentralized ledgers can help to protect LoRa data from being falsified when it is applied in LoRa systems. Therefore, a blockchain-enabled LoRa system with edge computing is expected to improve the performances of LoRa systems as well as the security. 
	
	\par
	There are few existed designs on blockchain enabled LoRa systems. Most of them can be roughly divided into two types. The former ones are to deploy blockchain only in the central cloud, e.g. in~\cite{blockchainlora, blockchainlora3}. Such a design can improve data security by the blockchain, but the edge computing abilities of LoRa gateways are not utilized. Meanwhile, the workload of central cloud is increased because central cloud needs to maintain the blockchain additionally. In the latter ones, the blockchain was proposed to directly be implemented at LoRa gateways without a specific design, which lacks of the implementation feasibility~\cite{blockchainlora2}. As the increasing data of the blockchain, the computing and storage resources in LoRa gateways might be exhausted soon. To the best of our knowledge, none of them has provided a feasible blockchain-based solution for LoRa with edge computing.

	\par
	Therefore, this paper aims to propose and implement a blockchain-enabled edge computing approach for LoRa systems with the open-source Hyperledger Fabric, which is called as HyperLoRa for the sake of illustration. A blockchain network with multiple ledgers is designed in HyperLoRa. Each node in blockchain network shares a whole copy of ledger that keeps transactions in chained blocks. Two ledgers are specially constructed for processing and storing different types of data, i.e., the delay tolerant application data with large size is in one ledger in the central cloud, while the time critical network data with small size is in another ledger at LoRa gateways. Then, the edge computing potentials of LoRa gateways can be well exploited. Two main features of the network servers, i.e., join procedure handling and application packages processing, can be migrated to LoRa gateways. As a consequence, the workloads of central cloud can be balanced by LoRa gateways, while the central processing unit (CPU) utilization in central cloud and the network bandwidth required between the LoRa gateways and the central cloud can be reduced. 
	\par
	Moreover, a HyperLoRa prototype is also implemented on embedded hardware to demonstrate the feasibility of our proposed design. In order to run Hyperledger Fabric on LoRa gateways that are built upon CPU of ARM64 architecture, the embedded Linux operating system is customized to let it fit with a third-party docker image of Fabric. Then, two channels are configured and governed by two organizations, i.e., LoRa gateway organization and network server organization. Each channel consists of an ordering node for consensus and many peers that maintain a shared ledger. The LoRa gateways can request and query transactions by invoking chaincodes via Fabric client. Moreover, the join server (JS) and network connector (NC) modules are implemented at LoRa gateways using the edge computing abilities. The prototype implementation of HyperLoRa demonstrates that it is viable to integrate blockchain into the LoRa gateways with the embedded hardware that are considered  resource-constrained. The experimental results show that HyperLoRa can have an equivalent performance but less resource consumption compared with traditional LoRa systems.

	\par
	The rest of the content is organized as follows. The related works are discussed in Section~\ref{sec:related}. Section~\ref{sec:overview} provides the overview of LoRa. The design of HyperLoRa system, as well as the edge computing functions of LoRa gateways are given in Section~\ref{sec:bclora}. Then, the security analysis is presented in Section~\ref{sec:security}. The implementation of HyperLoRa and the experiments with the results are elaborated in Section~\ref{sec:experiment}. Section~\ref{sec:conclusion} concludes the paper. 
	
	\section{Related works}
	\label{sec:related}
	Efficiency and security issues become critical for the development of IoT services. As a promising solution, blockchain has attracted lots of attention, since it can ensure the data security and privacy in a distributed manner~\cite{IoTBlockchain, iotblockc}, or provide a secure way to trade items~\cite{hyperledger}. Besides, edge computing can provide efficient and decentralized data services, such as data analytics or distributed machine learning at the edge, because it is closer the data source than central cloud~\cite{edgeai, fliotedge}. However, due to the various requirements and the limited resources in IoT devices, it is still challenging on how to integrate blockchain into different IoT scenarios both theoretically and technically. 
	\subsection{Open-source implementations of LoRa system}
	There are several implementations of LoRa system, e.g., the XisLoRa\footnote{https://github.com/xisiot/lora-system}, the ChirpStack\footnote{https://www.chirpstack.io/} and the Things Network (TTN)\footnote{https://www.thethingsnetwork.org/}. Their designs are all based on the reference model that is proposed in~\cite{lorawanbackend}. XisLoRa separates a network connection module to handle the package parsing and verification, while ChirpStack uses a gateway bridge to communicate directly with LoRa gateways. TTN follows the reference model and deploys one network server to handle all kinds of requests. These implementations have two things in common. On the one hand, all the LoRa gateways in these systems are only responsible for package forwarding. No further work is required for LoRa gateways to finish. Thus, the resources at LoRa gateways are not efficiently utilized. On the other hand, all the data in these systems are stored in traditional databases, such as MySQL (by XisLoRa), PostgreSQL (by ChirpStack) and InfluxDB (by TTN). Therefore, these systems have potential risks of data falsification and leakage without blockchain.
	\subsection{IoT system with blockchain}
	In traditional IoT systems, data are stored in centralized servers, which may cause eavesdropping and single-point bottleneck. The blockchain can be used to build a distributed and trustful data storage method and let all related entities to participate and supervise for data security. Among most of the literature, blockchain for IIoT becomes a widely studied scenario. Liang \emph{et al.}~\cite{blockchainiiot} proposed a blockchain based data transmission model for IIoT. By using secure cryptologic algorithm and decentralized consensus technique, the power data can be transmitted in a secure way. Wan \emph{et al.}~\cite{blockchainsmart} deploy a private blockchain in IIoT architecture in order to enhance the data security and privacy. The traditional storage layer is substituted by decentralized blockchain network. Moreover, Zhao \emph{et al.}~\cite{blockchainiiot2} propose a multi-chained IIoT platform to provide secure and decentralized data transmission between IIoT resources and end-users. Vehicular networks can also benefit a lot by blockchain. Kang \emph{et al.}~\cite{blockchainvehicular} deploy a Directed acyclic graph (DAG) based blockchain network in vehicular networks with edge computing. DAG is considered as a new kind of data structure for blockchain, in order to enhance the efficiency. However, it may cause inconsistency after some time. Jiang \emph{et al.}~\cite{blockchainiov} propose a distributed Internet of vehicles architecture with blockchain. The transmission delay is analyzed considering the mobility of vehicles. Blockchain can also be deployed in healthcare scenarios. In order to avoid the health data being eavesdropped by third-party servers, a blockchain based healthcare scheme is proposed where all end-devices, patients and doctors participate in the blockchain network~\cite{blockchainhealth}. Blockchain can also improve the progress of machine learning for IoT. Qiu \emph{et al.}~\cite{cloudedgeend} propose a collective $Q$-learning approach where the blockchain is used to share the training results between IoT nodes in order to govern the results in an efficient and secure way. 
	\subsection{Blockchain-enabled LoRa system with edge computing}
	As one of the promising IoT transmission techniques, LoRa is widely used for scenarios with low-power and wide-area requirements. Blockchain is suitable for a LoRa system to ensure data security. However, it is challenging to integrate blockchain into LoRa system. This is because most of the end-devices are resource-constrained. They cannot maintain ledgers in blockchain network. Meanwhile, the LoRa gateways are not as capable as central cloud, they are only used for transparent transmission of LoRa packages between end-devices and central cloud. Deploying blockchain in LoRa gateways needs a re-design of LoRa system architecture and well consideration of the resources. Liu~\emph{et al.}~\cite{loraedge} propose an edge computing based LoRa system in order to migrate the functions of rejoin and media access control (MAC) commands into LoRa gateways to mitigate the workloads of central cloud. However, databases are deployed in LoRa gateways to store data, which requires an extra mechanism for data synchronization. Both Lin~\emph{et al.}~\cite{blockchainlora3} and Danish~\emph{et al.}~\cite{blockchainlora} propose their designs for LoRa systems where blockchain is only deployed in central cloud. These designs cannot ease the burden of central cloud, but entails more workloads on it. The edge computing abilities of LoRa gateways are not enabled. Ozyilmaz~\emph{et al.}~\cite{blockchainlora2} propose a blockchain based LoRa system where entities with different computing and storage abilities play different roles. LoRa gateways with high abilities can download full blockchain, while weak LoRa gateways can only download the headers of blocks. However, such a design has ignored that the increasing size of the ledgers in the blockchain may exceed the storage limitation of LoRa gateways. Besides, the intention of deploying blockchain in LoRa gateways is to verify blocks and store data, making no benefit for the efficiency of the whole system.
	\par
	
	\section{Overview of LoRa system}
	\label{sec:overview}
	\subsection{\textbf{Network reference model}}
\begin{figure}[htb]
	\centering
	\includegraphics[width=0.4\textwidth]{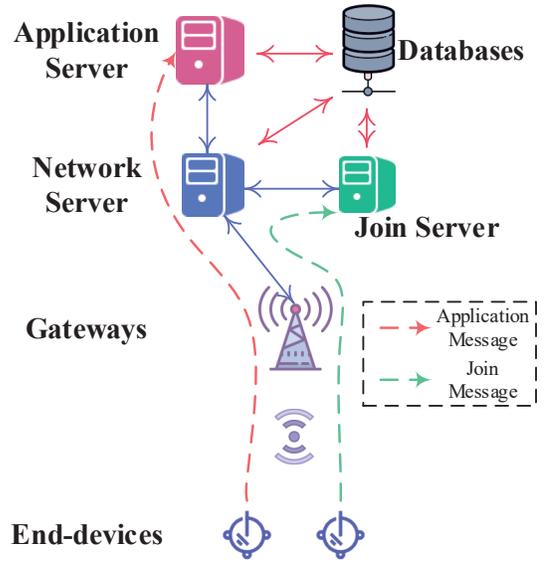}
	\caption{Illustration of LoRa network reference model.}
	\label{traditional}
	\end{figure}

	Most of the common LoRa systems are built based on the network reference model that is recommended in~\cite{lorawanbackend}. The specification suggests that the LoRa system should contain five basic parts, i.e., end-device, LoRa gateway, network server, join server and application server, which is illustrated in Fig.~\ref{traditional}. The end-devices are applied for IoT applications, such as sensing the environment. The application data are transmitted from end-devices to LoRa gateways via LoRa wireless links. The network servers, join servers and application servers are deployed in central cloud. LoRa gateways are responsible for package forwarding between end-devices and network servers. Then, all the application messages are processed by network servers. Among all these messages, the join messages are transferred to join servers, while the application messages are sent to application servers. All data are created, read, updated or deleted in databases.

	\subsection{\textbf{Basic features and categories of LoRa data}}
	The LoRa system mainly accomplishes two functions, i.e., the authentication of end-devices via join procedure and the transmission of application data between end-devices and application servers. 
	\begin{enumerate}
%		\item \textbf{Registry:} includes the registry of end-devices and gateways. Owners need to register the basic information to the LoRa network providers. 
		\item \textbf{Join procedure:} A join procedure is required for an end-device to be activated and access the LoRa network. There are two modes of join procedure, i.e., over-the-air activation (OTAA) and activation by personalization (ABP). OTAA works by dynamic negotiations between end-device and join server, while in ABP mode, all the session context are preset manually and remain unchanged both in end-device and join server. Either way, a session to exchange messages is established between end-device and network server.
		\item \textbf{Transmission of application data:} The application data are wrapped into LoRaWAN packages by end-devices for uplink transmission or by application servers for downlink transmission. Therefore, the network servers need to unpack/pack the packages, authenticate/specify the end-devices and verify/ensure the integrity of the packages for uplink/downlink transmission of application data.
%		\item \textbf{Network control:} LoRaWAN enables servers to control parameters of wireless transmission, e.g., the data rate or the transmission power. Control messages are encapsulated in MAC command fields, and encrypted by NwkSKey. 
	\end{enumerate}
	\par
	Basically, the data that are involved in the two functions can be classified into two categories, i.e., the session context data of end-devices and the application data. The context data contain all the necessary information to keep connections between end-devices and network servers. For example, the end-device address (DevAddr), the Application Session Key (AppSKey) for encryption/decryption of application data and the Network Session Key (NwkSKey) for calculation of message integrity code (MIC) value, and some configurations of LoRa physical layer. The application data are application-specific and owned by application providers or end-users. Take the environment monitoring as an example, the application data are the environment data that are gathered from end-devices with sensors. 

	\section{Design of HyperLoRa system}
	\subsection{\textbf{HyperLoRa architecture with multiple ledgers}}
	\label{sec:bclora}
\begin{figure}[htb]
		\centering
		\includegraphics[width=0.48\textwidth]{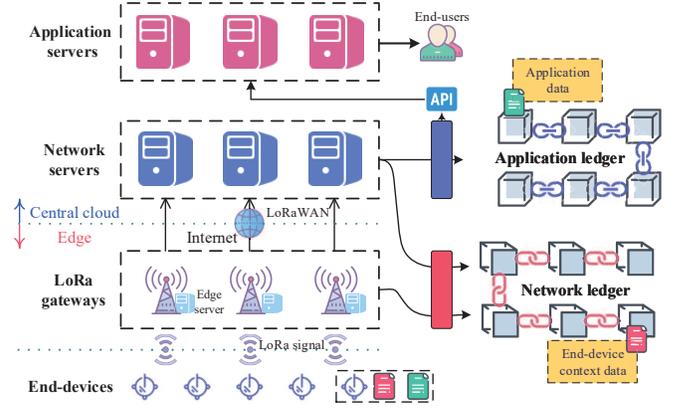}
		\caption{Illustration of HyperLoRa system architecture.}
		\label{architecture}
	\end{figure}

	The set of LoRa gateways and network servers are given as $\mathcal{G} = \{G_1, G_2, \dots, G_K\}$ and $\mathcal{S} = \{S_1, S_2, \dots, S_M\}$, respectively. $G_k$ stands for the GatewayEUI of the $k$-th gateway that is used to identify LoRa gateway, while $S_m$ represents the identity (ID) of network server $m$. Each LoRa gateway or network server has a pair of secret keys which are denoted as $<\pubkey, \prikey>$, where $\varepsilon\in\mathcal{G}\cup\mathcal{S}$. $\pubkey$ is the public key that is freely accessed by all other entities, while $\prikey$ is the private key that is only possessed by the owner. As illustrated in Fig.~\ref{architecture}, HyperLoRa consists of a network ledger $\mathcal{N}$ and an application ledger $\mathcal{A}$. All data are stored as immutable records of transactions in ledgers. The functions of the two ledgers are elaborated as follows, from perspectives of both data types and node types.
	\begin{itemize}
		\item \textbf{Perspective of data types:} All session context of end-devices are kept in $\networkc$. These data include all the attributes of end-devices that remain unchanged during one session. The Device Extended Unique Identifier (DevEUI) field is created when the end-device registers to join servers, while the other fields are generated during the join procedure. The application data that are produced by end-devices are stored in $\applicationc$. Only application servers are authorized to access these data for further processing. 
		
		\item \textbf{Perspective of node types:} $\networkc$ is maintained by all LoRa gateways and network servers. LoRa gateways can generate new blocks for $\networkc$ in OTAA mode, while network servers create new blocks for end-devices in ABP mode. This is because LoRa gateways receive join requests from OTAA-mode end-devices directly. Thus, they can pack session context data into blocks without wasting time and resources forwarding data to network servers. Both LoRa gateways and network servers are responsible for block verification and synchronizing on their ledgers. As for $\applicationc$, the new blocks are generated and validated only by network servers. However, some application programming interfaces (APIs) are provided for the development of application servers.
	\end{itemize}
	\par
	The session context data and the application data have different characteristics and purposes. The session context data are used to authenticate end-devices, encrypt messages and calculate MIC. The amount of session context data is small and the changing frequency is low. Therefore, LoRa gateways can be able to maintain $\networkc$, because there are extra resources apart from those that are used by physical layer. Different from that, the size of $\applicationc$ can be increasing quickly. Besides, application data should only be accessed by application servers to provide data services for end-users. Thus, only network servers can maintain $\applicationc$. Although all the data are generated by end-devices originally, they cannot participate in blockchain network due to their limitation on resources.
	\par
	
	\begin{table*}[htb]
		\centering
		\caption{List of mathematical notations}
		\label{tab:notation}
		\begin{tabular}{c|c|c|c}
			\firsthline
			\textbf{Notation} & \textbf{Explanation} & \textbf{Notation} & \textbf{Explanation} \\
			\hline
			$\mathcal{G}$ & Set of LoRa gateways & $\mathcal{S}$ & Set of network server \\
			\hline
			$\varepsilon$ & An entity of LoRa gateway or network server & $\pubkey$ & Public key of entity $\varepsilon$ \\
			\hline
			$\prikey$ & Private key of entity $\varepsilon$ & $\networkc$ & Network ledger \\
			\hline
			$\applicationc$ & Application ledger & $B$ & Block \\
			\hline
			$T$ & Transaction & $h$ & Message digest of transaction \\
			\hline
			$t$ & Request time of transaction & $d$ & Raw data in transaction \\
			\hline
			$\tau$ & Generation timestamp of block & $\mathcal{M}$ & Root value of Merkle tree \\
			\hline
			$\mathcal{H}$ & Hash value of previous block & & \\
			\lasthline
		\end{tabular}
	\end{table*}
	\subsubsection{\textbf{Network ledger}}
	The context data are recorded in $\networkc$ in the form of transactions which is denoted as $T^\networkc$. The content of $T^\networkc$ is given as,
	\begin{equation}
	T^\networkc = <\varepsilon, h, t, \bar{d}>,
	\end{equation}
	where $h$ is the message digest of the transaction, $t$ is the timestamp of the transaction and $\bar{d}$ is the session context data after encryption. The raw $d$ is given as,
	\begin{equation*}
	d = \text{ DevEUI } || \text{ AppKey } || \text{ DevAddr } || \text{ NwkSKey } || \text{ Nonces },
	\end{equation*}
	where operator $||$ means the concatenation of the data, and $\text{Nonces} = \text{DevNonce} || \text{AppNonce}$. Note that the AppSKey is not included because it is related to applications and should only be kept by application servers. Each LoRa gateway needs to encrypt $d$ and calculate $h$ to generate a $T^\networkc$ using the key pairs. $\bar{d}$ is calculated as,
	\begin{equation}
	\label{eq:enc}
	\bar{d} = \text{Enc}\{\pubkey, d\}
	\end{equation}
	and $h$ is derived as,
	\begin{equation}
	h = \text{Sign}\{\prikey, t || \bar{d}\},
	\end{equation}
	where operator $\text{Enc}$ stands for an encryption algorithm using the public key of the requester and $\text{Sign}$ represents a digital signature algorithm with private key. Therefore, $d$ can only be decrypted out using the corresponding private key, and $h$ can be verified by any LoRa gateway or network server that has the public key of the requester.
	\par
%	only gateways can generate new blocks. Servers serve as endorsers that can verify the blocks. Gateways keep collecting join requests and turn them into transactions.
	In $\networkc$, after reaching a pre-defined condition (e.g. a period of time or maximum number of transactions), the LoRa gateway starts to generate a new block that contains all the transactions after that period. Assuming that totally $R$ transactions need to be packaged at time $\tau$, and letting $T_r^\networkc = <\varepsilon_r, h_r, t_r, \bar{d}_r>$ denotes the $r$-th transaction, the LoRa gateway needs to construct a Merkle tree before generate the block according to Algorithm~\ref{alg:merkletree}.
	\begin{algorithm}[!htb]
		\KwIn{All transactions $\{T_1^\networkc, T_2^\networkc, \dots, T_R^c\}$, let $R = 2^x + y$, where $x \geq 0, 0 \leq y < 2^x$ and $z = 1$}
		\KwOut{The root hash $\mathcal{M}$}
		\BlankLine
		
		\For{$z$ in $1, 2, \dots, x + 1$}{
			\For{$i$ in $1, 2, \dots, \ceil{\frac{R}{2^z}}$}{
				\If{$z = 1$}{
					\If{$2i \leq R$}{
						\begin{equation*}
							hash_{z, i} = H(h_{2i - 1} || h_{2i}),
						\end{equation*}
						where $H$ is a hash function.
					}
					\Else{
						\begin{equation*}
							hash_{z, i} = h_{2i - 1}
						\end{equation*}
					}
				}
				\Else{
					\If{$2i \leq \ceil{\frac{R}{2^{z - 1}}}$}{
						\begin{equation*}
							hash_{z, i} = H(hash_{z - 1, 2i - 1} || hash_{z - 1, 2i}),
						\end{equation*}
					}
					\Else{
						\begin{equation*}
							hash_{z, i} = hash_{z - 1, 2i - 1}
						\end{equation*}					
					}
				}
			}
		}
		$\mathcal{M} = hash_{x + 1, 1}$.
		\caption{The generation of Merkle tree and root value.}
		\label{alg:merkletree}
	\end{algorithm}
	\par
	After creating the Merkle tree and getting the root value $\mathcal{M}$, LoRa gateway starts to generate a block $B^\networkc$ which includes an index indicator $\zeta\in \mathbb{Z}$, a body $b^\networkc$ and a header $e^\networkc$. $b^\networkc$ is given as follows,
	\begin{equation*}
		b^\networkc = \varepsilon_1||h_1 || t_1 || \bar{d}_1 ||\varepsilon_2 || h_2 || t_2 || \bar{d}_2 || \dots || \varepsilon_R || h_R || t_R || \bar{d}_R
	\end{equation*}
	Keeping the $\varepsilon$ field in blocks are necessary, because transactions can be transferred from one LoRa gateway to others. The $e^\networkc$ is given as $e^c = <\tau, \mathcal{M}, \mathcal{H}>$, where $\tau$ is the timestamp of the generation of the block, $\mathcal{M}$ is the root of Merkle tree and $\mathcal{H}$ is the hash value of the previous block. For the $i$-th block, the $\mathcal{H}_i$ is derived as follow,
	\begin{equation*}
		\mathcal{H}_i = H(\zeta_{i - 1} || e_{i - 1}^\networkc || b_{i - 1}^\networkc)
	\end{equation*}
	\par
	The new block is broadcast to all other LoRa gateways and network servers in $\networkc$. Others need to verify the block and endorse it if it is considered valid. At first, other nodes need to check $h$ of all $T^\networkc$ to make sure they come from exactly a trustful LoRa gateway. With the ID value, the $\pubkey$ can be derived, and $h$ can be verified using the same algorithm as the signing procedure. Then, the root of Merkle tree is calculated again to ensure its validity. Note that the session context data of end-devices are encrypted, thus, the data cannot be eavesdropped during transmission, and other nodes need not to verify the raw content of transactions. 
	\par
	After verification, other nodes need to broadcast a signed result. According to practical Byzantine fault tolerance (PBFT), if each node receives more than $2p + 1$ (where $p$ is the maximum possible node that is not functioning well) results that are the same, then the consensus is reached. All nodes synchronize $B^\networkc$ on their ledger.
	\subsubsection{\textbf{Application ledger}}
	The application data are stored in $\applicationc$ as transactions which is denoted as $T^\applicationc$. The structure of $T^\applicationc$ is similar to $T^\networkc$ except for the identifier of requester and the data field. Let $T^\applicationc = <\varepsilon, h, t, d>$, where $\varepsilon\in\mathcal{S}$ and $d$ is the application data. Note that the requester of $T^\applicationc$ can only be network servers, and the data in transactions need not to be deciphered by network servers, because the application data have been encrypted by end-devices using AppSKey and kept encrypted during the processing. 
	\par
	When a network server receives enough $T^\applicationc$, it starts to wrap them into a new block. By reaching consensus with all other network servers, the new block can be recorded on $\applicationc$, and the application data $d$ is finally stored on the shared ledger. End-users can access the application data through application servers that could get application data from APIs and decrypt them using proper AppSKey.
	\par
	\begin{table*}[htb]
		\centering
		\caption{Information of each component in HyperLoRa system}
		\label{tab:components}
		\begin{tabular}{c|c|c|c}
			\firsthline
			\textbf{Component} & \textbf{Functions} & \makecell{\textbf{Capability of} \\ \textbf{computing and storage}} & \textbf{Ledger} \\
			\hline
			End-device & \makecell{Sensing, actuating, $\dots$} & Very low & None \\
			\hline
			LoRa gateway & \makecell{Signal processing, data forwarding, \\join procedure handling, application packages processing } & Mediocre & $\networkc$ \\
			\hline
			Network server & \makecell{Data analytics, application APIs, $\dots$} & High & $\networkc, \applicationc$ \\
			\lasthline
		\end{tabular}
	\end{table*}
	
	\subsection{\textbf{Edge computing in LoRa gateways}}
	With edge computing abilities in HyperLoRa, two modules are moved from network servers to LoRa gateways, i.e., JS that handles the join procedures of end-devices and NC that is responsible for application package processing. By join procedures, the context data of end-devices are generated and stored in $\networkc$. On the other hand, LoRa gateways need these context data from $\networkc$ to fulfill the tasks of application package processing.
	\par

		\begin{figure}[htb]
		\centering
		\includegraphics[width=0.48\textwidth]{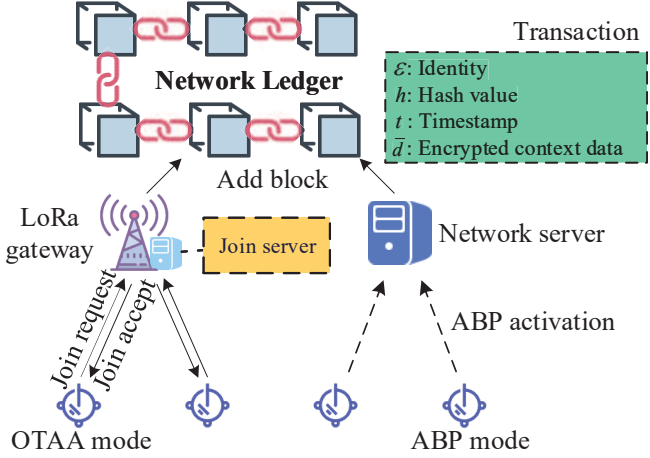}
		\caption{Illustration of join procedure.}
		\label{join}
	\end{figure}
	\begin{figure}[htb]
		\centering
		\includegraphics[width=0.48\textwidth]{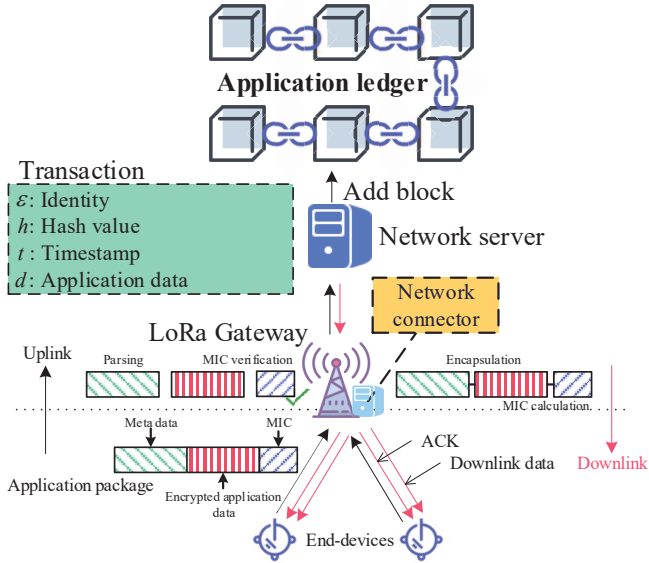}
		\caption{Illustration of application package processing.}
		\label{app}
	\end{figure}

	\subsubsection{\textbf{Join procedure handling}}
	As shown in Fig.~\ref{join}, the join procedure of OTAA mode is mainly handled by LoRa gateways, while the ABP mode activation of end-devices are processed by network servers. 

\begin{figure}[htb]
		\centering
		\includegraphics[width=0.48\textwidth]{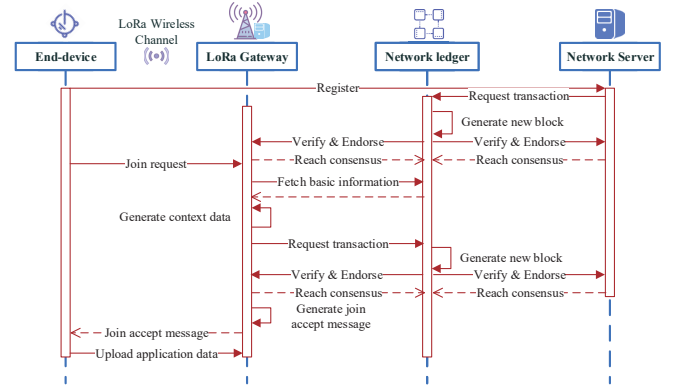}
		\caption{Workflow of join procedure in OTAA mode.}
		\label{joinprocedure}
	\end{figure}

	\begin{itemize}
		\item \textbf{OTAA:} As shown in Fig.~\ref{joinprocedure}, when a LoRa gateway receives a join request, it checks the validity of the request, including the MIC value. Then, LoRa gateway needs to generate session context data for this end-device, including the DevAddr, two session keys and some meta data. A transaction that contains these data is requested and put into a new block by the LoRa gateway. After consensus, the new block is added into $\networkc$. In the meantime, the LoRa gateway has to generate a join accept message and send it back to end-device. Successful reception of join accept message by end-device indicates the success of the join procedure. 
		\item \textbf{ABP:} In ABP mode, the manufacturers or owners of end-devices need to fill all session context data manually via the interfaces offered by network servers. Then, network servers request transactions of these data and add blocks on $\networkc$ after consensus. No more negotiations are needed between end-devices and JS in LoRa gateways before end-devices can upload application data. 
	\end{itemize}
	\par
	
	\subsubsection{\textbf{Application packages processing}}
\begin{figure*}[htb]
		\centering
		\includegraphics[width=0.8\textwidth]{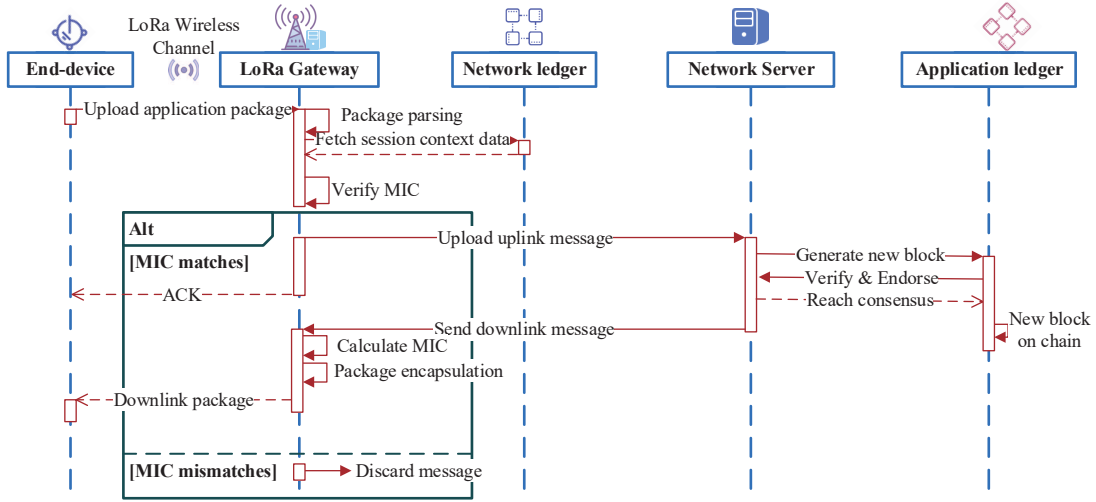}
		\caption{Workflow of application package processing.}
		\label{verification}
	\end{figure*}

	Owing to $\networkc$ being deployed in LoRa gateways, an NC module that is used for application package processing is moved from network server to LoRa gateway, as shown in Fig.~\ref{app}. The workflow of message exchange between end-device and the HyperLoRa system is depicted in Fig.~\ref{verification}.

	%The application data, as well as the status information of both end-device and gateway, are transmitted to server. In order to save the application data, server issues a transaction to the application blockchain. The verification and consensus are reached between all the other servers. Then, a new block is added. 
	\par
	\begin{itemize}
		\item \textbf{Uplink package:}
		When the end-device uploads an application package, the NC in LoRa gateway parses the package into three parts, i.e, the meta data, the encrypted application data and the MIC value. By extracting the DevAddr field from the meta data, NC can query the session context data of the end-device from $\networkc$. Then, it checks the validity of MIC value to ensure the integrity of the package. If the MIC value mismatches, the package is discarded directly. Otherwise, LoRa gateway uploads the encrypted application data to network server. The application data are packed into blocks that will be added to $\applicationc$ by network servers. In the meantime, LoRa gateway needs to inform end-device of the successful reception of package with an acknowledgment (ACK) message. 
		\item \textbf{Downlink package:}
		If there are any downlink data for end-devices, network servers need to reply the downlink data to LoRa gateways at the same time as adding new block on chain. Only encrypted data need to be sent from network server to LoRa gateways rather than full packages. Then, the downlink package including the information and the MIC value that are created and calculated by LoRa gateways are encapsulated in NC. During this process, the context data of end-devices are retrieved by $\networkc$. The downlink package is finally transmitted to end-devices via LoRa wireless channels.
	\end{itemize}
	\par
 	With edge computing that uses the LoRa gateway near the end-devices to perform preliminary processing, the central cloud efficiency of HyperLoRa can be optimized. The works of package parsing, encapsulation and MIC calculation are finished by LoRa gateways that are on the edge rather than network servers that are on the central cloud. Thus, a lot of computing resources of central cloud are saved, while the spare resources from LoRa gateways are utilized. Besides, the invalid messages such as those that have wrong MIC values are blocked by LoRa gateways without affecting both the computing resources in central cloud and the transmission links between LoRa gateways and central cloud. 

	\section{Security analysis}
	\label{sec:security}
	In this section, the improvement of security brought by HyperLoRa system is analyzed with respect to four kinds of risks, i.e, application-layer Denial of Service (DoS) attack, Single Point Of Failure (SPOF), malicious LoRa gateway and malicious network server. These risks can be mitigated or eliminated by the deployment of blockchain and edge computing in HyperLoRa.
	\subsection{\textbf{Application-layer DoS attack}}
	For traditional LoRa system, the interfaces between network servers and LoRa gateways is defined in~\cite{semtech}. Therefore, an attacker can launch DoS attack by sending a large number of invalid application packages to network servers in high frequency to exhaust computing and bandwidth resources. In HyperLoRa system, the application packages are processed in LoRa gateways, and the interface protocol between network servers and LoRa gateways can be defined privately. It could take attackers much more effort to crack the private protocol and launch DoS attack towards network servers. It is also very hard to attack HyperLoRa through LoRa gateways, because they are geographically distributed and they only receive physical LoRa signals. LoRa gateways can also filter malicious traffic for network servers, thus, it is guaranteed that network servers can stay healthy to ensure the functions of the HyperLoRa system during the DoS attack through LoRa gateways.
	\subsection{\textbf{SPOF}}
	In HyperLoRa, the SPOF of both LoRa gateways and network servers can be solved, because each LoRa gateway or network server maintains a full copy of the shared ledger. When one LoRa gateway malfunctions, the nearby LoRa gateways can take the place as long as they can receive the LoRa signals from end-devices. They only need to get the private key of the broken LoRa gateway in a secure way. When one network server breakdowns, other network servers can replace it, while LoRa gateways can re-uploads data to the alternative network servers. On the one hand, the JS and NC are migrated into LoRa gateways, so that the malfunctions of one network server has no effect on the handling of join procedures and application package processing. On the other hand, some of the workloads of network servers are balanced by LoRa gateways with edge computing, thus, the crashing possibility of network servers is decreased a lot.
	\subsection{\textbf{Malicious LoRa gateway}}
	A malicious LoRa gateway can do two things to HyperLoRa system. The first kind is to steal session context data of end-devices from $\networkc$. The other kind of thing is to disturb the consensus process. In both ways, the malicious LoRa gateway needs to pass the authentication process and participate in the blockchain in HyperLoRa. If the shared ledger is retrieved, the malicious LoRa gateway also needs to obtain the $k^{\text{pri}}$ of other LoRa gateways to decrypt the session context data in blocks. This could be extremely difficult because LoRa gateways can keep the $k^{\text{pri}}$ in specialized encryption chips. For consensus process, attackers need to crack more than $p$ LoRa gateways in HyperLoRa to break the PBFT algorithm, which is difficult to realize. 
	\subsection{\textbf{Malicious network server}}
	Assume an attacker has deployed in HyperLoRa system a malicious network server that has successfully passed the authentication and has established connections with some LoRa gateways (which is actually difficult to realize). Then, the malicious network server can share both $\networkc$ and $\applicationc$. However, neither the session context data of end-devices nor the application data can be stolen, falsified or destroyed. First of all, all the session context data are encrypted using $k^{\text{pri}}$, while all the application data are encrypted using AppSKey that is kept only by end-devices and application servers. To decipher the data, the attacker needs also to get all these keys. Then, the blockchain in HyperLoRa ensures that the data inside the blocks cannot be altered. Moreover, all nodes share a full copy of ledgers, thus, deleting data from one network server will not damage the data that are kept in other LoRa gateways or network servers. 
	
	\section{Implementation and Experiments}
	\label{sec:experiment}
	\subsection{\textbf{Implementation with Hyperledger Fabric}}
\begin{figure}[htb]
		\centering
		\includegraphics[width=0.49\textwidth]{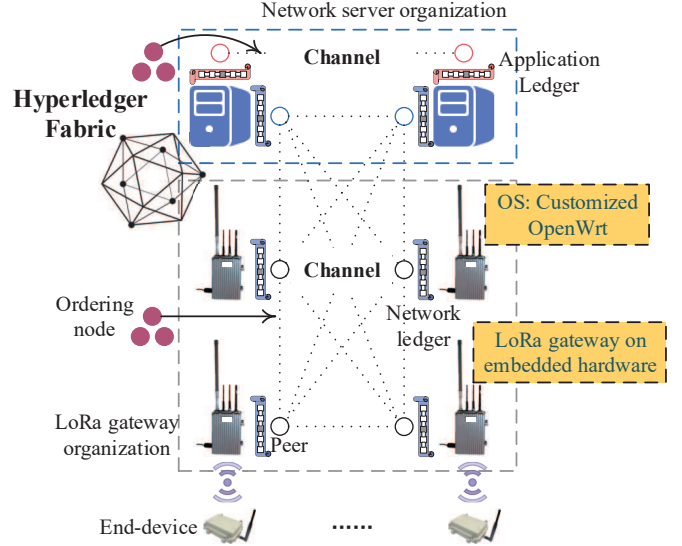}
		\caption{Implementation of HyperLoRa with embedded LoRa gateways and Hyperledger Fabric.}
		\label{implementation}
	\end{figure}

	The prototype of HyperLoRa system is implemented where the blockchain is realized using Hyperledger Fabric, as shown in Fig.~\ref{implementation}. The end-device is deployed with a LoRa transceiver module SX1276 from Semtech. By plugging in proper sensor, the end-device can collect data and upload to LoRa gateway. The LoRa gateway is developed on embedded hardware with Smart6818 CPU board that is industrial specialized and uses Samsung S5P6818 octa-core Cortex-A53 System on Chip (SoC). The edge computing can be achieved with such hardware. Besides, two radio frequency front-ends (SX1255) are deployed and connected to a digital baseband chip SX1301 for signal processing. 
	\par
	In order to deploy Fabric on LoRa gateway, the operating system is customized based on Linux OpenWrt project since Fabric requires cgroup feature and overlay file system. In the meantime, a third-party docker image of Fabric is built to run on the LoRa gateway. During the initialization of Fabric, two organizations are created, i.e., LoRa gateway organization that contains peers that are deployed in LoRa gateways and network server organization that includes peers that are deployed in network servers. Each peer acts as both endorser and committer. Two channels are then established, each of which maintains ledger $\networkc$ and $\applicationc$. All peers in two organizations can communicate in the channel of $\networkc$, while only peers in network server organization can communicate secretly in channel of $\applicationc$. As a permissioned blockchain, a Membership Service Provider (MSP) is built to handle the authentications and authorizations of all peers. That is to say, only authorized LoRa gateway or network server can join in the blockchain network. The consensus is accomplished using PBFT algorithm by several ordering nodes in each channel. Due to the limited computational capabilities of LoRa gateways, all ordering nodes in two channels are deployed in the central cloud. 
	\par
	Several chaincodes are developed and hosted in LoRa gateways peers to manipulate data in $\networkc$. Both JS and NC can invoke the chaincodes by Fabric client to create, read or delete session context data of end-devices. Then, peers try to run the chaincodes in a virtual environment to ensure they are valid. By receiving the responses from peers, JS or NC sends the transaction to the ordering nodes for consensus. After consensus, the data are recorded on all distributed $\networkc$ and the latest values of specific data fields are kept as the world states of $\networkc$ in a state database.
	\par
	\subsection{\textbf{Experimental setup}}
	Our prototype system consists of $4$ LoRa gateways and $2$ network servers at the central cloud. To emulate hundreds of end-devices, several customized Locust\footnote{https://www.locust.io/} clients is deployed on independent servers that can connect to LoRa gateways directly. A Fabric network of version $1.4.0$ is established on both LoRa gateways and network servers. The hardware information is summarized in Table~\ref{tab:hardware}.
	\begin{table*}[htb]
		\centering
		\caption{Hardware information}
		\label{tab:hardware}
		\begin{tabular}{c|c|c|c}
			\firsthline
			\textbf{Type of device} & \textbf{Processor} & \textbf{Number of CPU cores} & \textbf{Memory} \\
			\hline
			LoRa gateway & Samsung(R) S5P6818 @ \SI{1.40}{\GHz} & $8$ & \SI{1}{\giga\byte}\\
			\hline
			Network server & Intel(R) Xeon(R) W-2123 @ \SI{3.60}{\GHz} & $8$ & \SI{32}{\giga\byte} \\
			\hline
			Locust client server & Intel(R) Xeon(R) CPU E5-2609 v4 @ \SI{1.70}{\GHz} & $8$ & \SI{16}{\giga\byte}\\
			\lasthline
		\end{tabular}
	\end{table*}
	\par
	Some of the key experimental parameters are listed in Table~\ref{tab:parameters}. The ordering type is \textit{solo}, which means the consensus is reached by a single ordering node using sorting algorithm. New block is generated every \SI{2}{\second}. If the total number of transactions that are received within \SI{2}{\second} exceeds $200$, a new block is also forced to be created. 
	\begin{table}[htb]
		\centering
		\caption{Parameter settings}
		\label{tab:parameters}
		\begin{tabular}{c|c|c}
			\firsthline
			\textbf{Name} & \textbf{Value} & \textbf{Description} \\
			\hline
			\multicolumn{3}{c}{\textbf{General settings}}\\
			\hline
			OrdererType & Solo & Type of order in Hyperledger \\
			\hline
			BatchTimeout & \SI{2}{\second} & Time interval to generate a block \\
			\hline
			MaxMessageCount & $200$ & Maximum number of one batch \\
			\hline
			\multicolumn{3}{c}{\textbf{Join requests}} \\
			\hline
			MaxPackageInterval & \SI{2}{\hour} & \makecell{Maximum intervals between two \\packages of one end-device} \\
			\hline
			MinPackageInterval & \SI{10}{\minute} & \makecell{Minimum intervals between two \\packages of one end-device} \\
			\hline
			Timeout & \SI{300}{\second} & Timeout of each join request \\
			\hline
			\multicolumn{3}{c}{\makecell{\textbf{Application packages by authorized end-devices} \& \\\textbf{Application packages by authorized and unauthorized end-devices}}}\\
			\hline
			MaxPackageInterval & \SI{17}{\second} & - \\
			\hline
			MinPackageInterval & \SI{13}{\second} & - \\
			\hline
			Timeout & \SI{30}{\second} & Timeout of each application package\\
			\lasthline
		\end{tabular}
	\end{table}

	\par
\begin{figure}[htb]
		\begin{subfigure}[b]{.24\textwidth}
			\centering
			\includegraphics[width=0.99\textwidth]{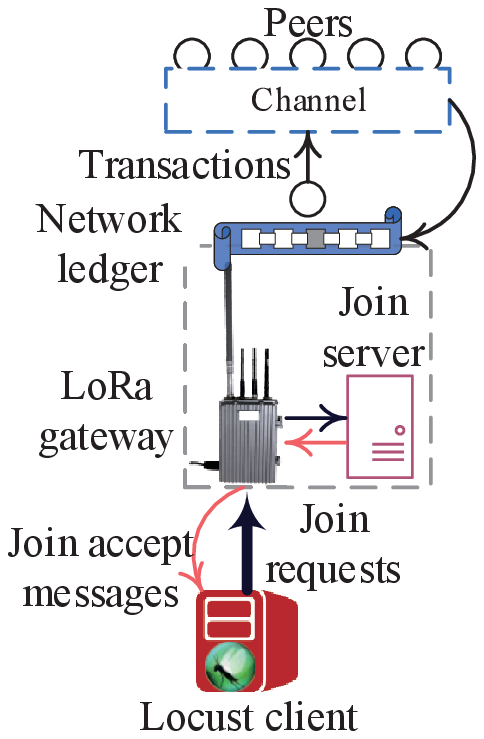}
			\caption{\textit{Experiment 1}: join requests.}
			\label{exp1}
		\end{subfigure}
		\begin{subfigure}[b]{.24\textwidth}
			\centering
			\includegraphics[width=0.99\textwidth]{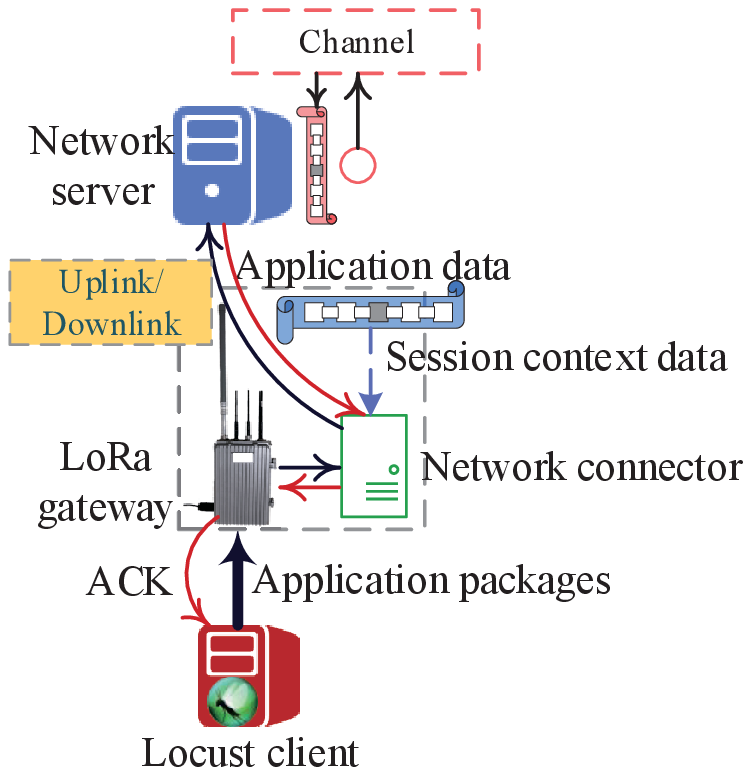}
			\caption{\textit{Experiment 2}: application packages by authorized end-devices.}
			\label{exp2}
		\end{subfigure}
		\newline
		\begin{subfigure}[b]{0.24\textwidth}
			\centering
			\includegraphics[width=0.99\textwidth]{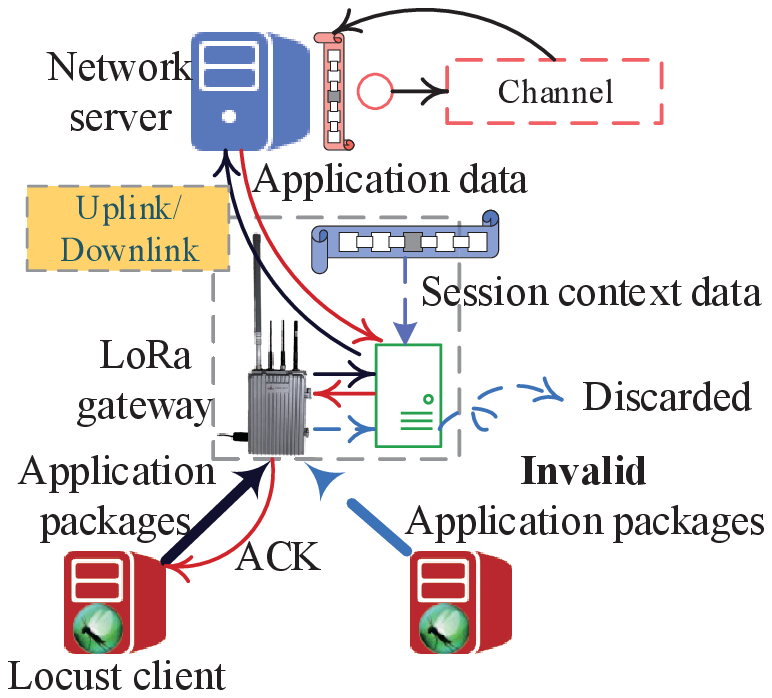}
			\caption{\textit{Experiment 3}: application packages by authorized and unauthorized end-devices.}
			\label{exp3}
		\end{subfigure}
		\begin{subfigure}[b]{0.24\textwidth}
			\centering
			\includegraphics[width=0.99\textwidth]{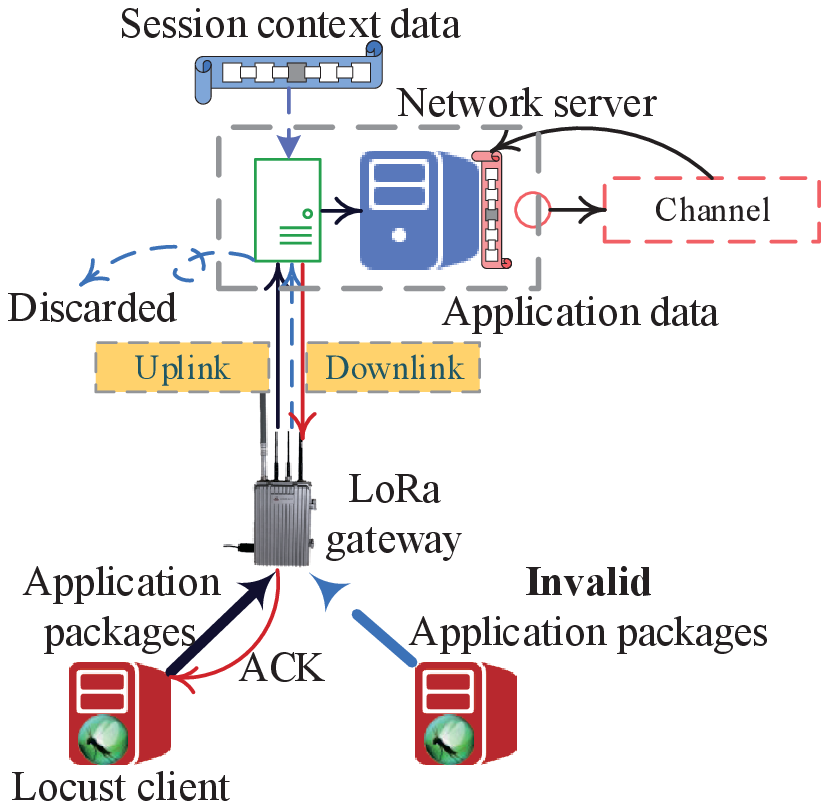}
			\caption{Comparison experiment with traditional LoRa system in \textit{Experiment 3}.}
			\label{exp4}
		\end{subfigure}
		\caption{Illustrations of the experiments on HyperLoRa prototype.}
		\label{experiment}
	\end{figure}

	Three kinds of experiments are conducted, i.e.,
	\begin{enumerate}
		\item \label{enum:1}\textbf{Experiment 1 (join requests):} 
		$100\sim2,200$ end-devices are emulated to continuously send join requests to LoRa gateways. Each LoRa gateway manages $25\sim 550$ end-devices. As shown in Fig.~\ref{experiment}(\subref{exp1}), each end-device runs periodically with a randomly chosen interval between \SI{10}{\minute} and \SI{2}{\hour}. If a join accept message is received by end-device, the join procedure is considered successful. If no apply is received after \SI{300}{\second}, the join request is failed. The detailed processing time of join requests are recorded. 
		\item \label{enum:2}\textbf{Experiment 2 (application packages by authorized end-devices):} 
		$100\sim2,000$ end-devices are emulated in this experiment. Each LoRa gateway covers $25\sim400$ end-devices equivalently. The join procedures of all end-devices are accomplished. Thus, end-devices are authorized to send application packages. Each end-device randomly chooses an interval between \SI{13}{\second} and \SI{17}{\second} to upload packages constantly. The network server replies each package with an acknowledgment to indicate that the uplink data are received successfully, as shown in Fig.~\ref{experiment}(\subref{exp2}). If no acknowledgment is received by end-device for \SI{30}{\second} after it uploads, the transmission of application package is considered failed. End-device keeps sending after the time interval or any failure occurs. By running this experiment, the maximum possible performance of the prototype can be evaluated. The processing time, CPU utilization and system throughput are recorded. 
		\item \label{enum:3}\textbf{Experiment 3 (application packages by authorized and unauthorized end-devices):} 
		In this experiment, $100\sim2,000$ end-devices are emulated in total. $4$ LoRa gateways share equal number of end-devices. Among all the end-devices under the coverage of each LoRa gateway, half of them have finished the join procedures, while the other half are considered unauthorized to the prototype. All the end-devices send application packages to LoRa gateways in the same way as the \textit{Experiment 2}. For those unauthorized end-devices, their packages cannot pass the MIC verification process in LoRa gateways and will be discarded directly, as shown in Fig.~\ref{experiment}(\subref{exp3}). This experiment can simulate the situation that the system is under the application-layer DoS attack. The system throughput, CPU utilization and network bandwidth occupation are recorded. Moreover, a comparison experiment with traditional LoRa system approach is conducted. The application packages are sent to a traditional LoRa system where both the JS and NC are implemented in central cloud rather than in the LoRa gateways. Therefore, central cloud needs to process all valid and invalid application packages. A blockchain network is also established in the traditional LoRa system, but both $\networkc$ and $\applicationc$ are maintained only by network servers, as illustrated in Fig.~\ref{experiment}(\subref{exp4}). The comparison can show the improvement of performance brought by HyperLoRa.
	\end{enumerate}

	\begin{figure*}[!t]
		\centering
		\includegraphics[width=0.65\textwidth]{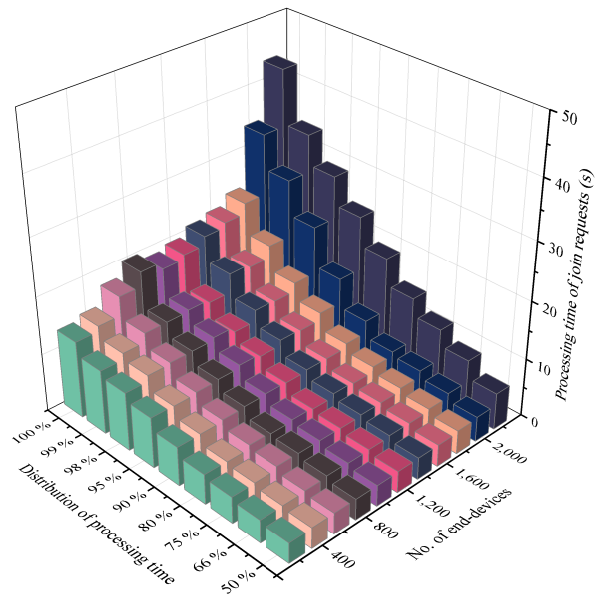}
		\caption{Distribution of processing time of join requests in \textit{Experiment \ref{enum:1}}.}
		\label{joindistribution}
	\end{figure*}
	
	\begin{figure}[!h]
		\centering
		\includegraphics[width=0.45\textwidth]{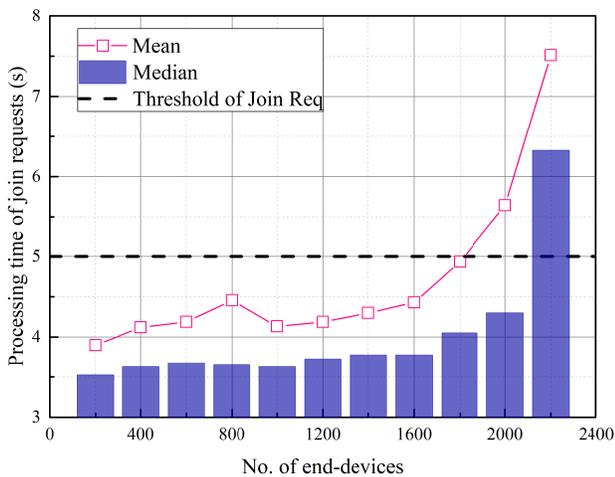}
		\caption{Processing time statistics of join requests in \textit{Experiment \ref{enum:1}}.}
		\label{joinlatency}
	\end{figure}
	\begin{figure}[!h]
		\centering
		\includegraphics[width=0.45\textwidth]{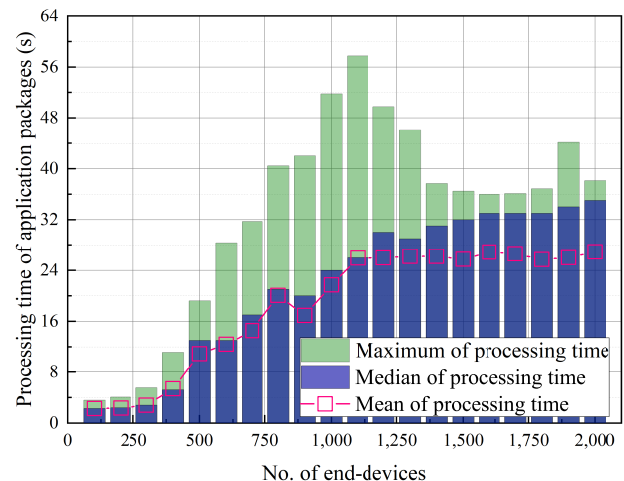}
		\caption{Illustration of processing time of application packages in \textit{Experiment \ref{enum:2}}.}
		\label{applatency}
	\end{figure}

	\subsection{\textbf{Results and analysis}}
	Figure~\ref{joindistribution} shows the detailed distribution of processing time of join requests. Each bar represents the percentage of end-devices whose processing time locates within the value of the top point for specific number of end-devices. The processing time rises slightly with the increase of the number of end-devices before $2,000$. When the number of end-devices comes to around $500$ per LoRa gateway, the processing time has a significant growth. The same trend is shown in Fig.~\ref{joinlatency}. It presents the statistical results of processing time of join requests. The black-dash line is the maximum allowable latency during which an end-device can receive the join accept message. It can be configured according to specific requirements (\SI{5}{\second} in the \textit{Experiment 1}). The join request is considered failed if time exceeds the threshold. The mean latency starts to exceed the threshold after the number of end-devices passes $2,000$. The results demonstrate the maximum capability of handing join requests in this system, i.e., for \SI{5}{\second} restriction and $1,800$ end-devices, nearly $75\%$ can receive join accept messages successfully.
	\par
	Figure~\ref{applatency} illustrates the statistical processing time of application packages in \textit{Experiment 2}. The mean latency keeps rising as the number of end-devices increases before $1,100$, because more end-devices generate more application data, costing more time to process. When the number exceeds $1,200$, the processing time becomes stable, this is because the HyperLoRa system has reached its maximum load. There are no further resources to handle more application packages. Then, these packages are considered being discarded. This result shows that the prototype can handle totally $1,200$ end-devices whose uploading intervals are around \SI{15}{\second} and timeouts are \SI{30}{\second}. Note that the uploading frequency is far more than most of the sensing based IoT applications. In practical, the maximum number can still be increasing.

	\begin{figure}[!h]
		\centering
		\includegraphics[width=0.45\textwidth]{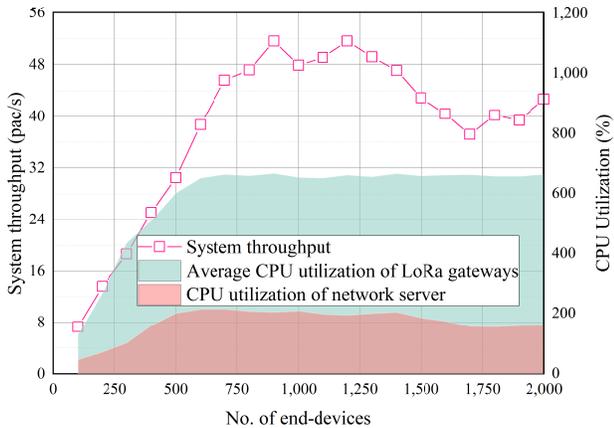}
		\caption{Illustration of system throughput of application packages, as well as CPU utilization of LoRa gateways and network server in \textit{Experiment \ref{enum:2}}.}
		\label{rpsandcpu}
	\end{figure}

	\begin{figure}[!h]
		\centering
		\includegraphics[width=0.45\textwidth]{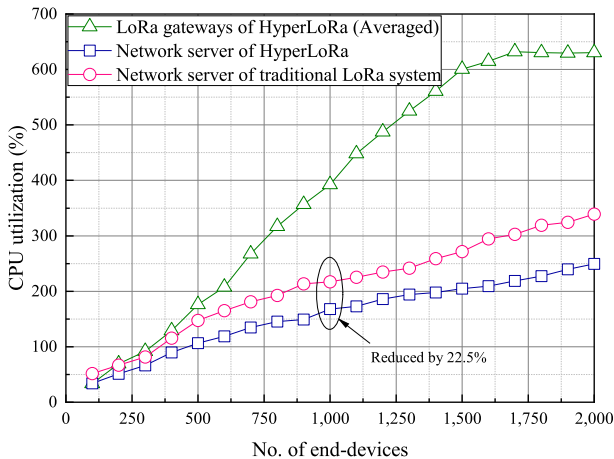}
		\caption{Illustration of CPU utilization of LoRa gateways and network server with both valid and invalid application packages in \textit{Experiment \ref{enum:3}}.}
		\label{cpumix}
	\end{figure}
	\par
	During the processing of valid application packages, the system throughput and the CPU utilization of both LoRa gateways and network server are shown in Fig.~\ref{rpsandcpu}. Before $1,100$ end-devices, the throughput keeps rising, because the resources are not fully utilized. In the meantime, the CPU utilization is increasing. After the number of end-devices exceeds $1,200$, the throughput becomes stable, and even declines a little. Similar to the trend that is presented in Fig.~\ref{applatency}, it is because the system is in full-load condition. This is also reflected by the change of the CPU utilization. Nearly $7$ cores of CPU in LoRa gateways are occupied by NC module. The rest core is taken up by the operation systems and the signal processing modules. However, the CPU utilization of network server remains in a low level, because LoRa gateways with edge computing have taken the works of packages processing and verification. 
	\par

\begin{figure}[!h]
	\centering
	\includegraphics[width=0.45\textwidth]{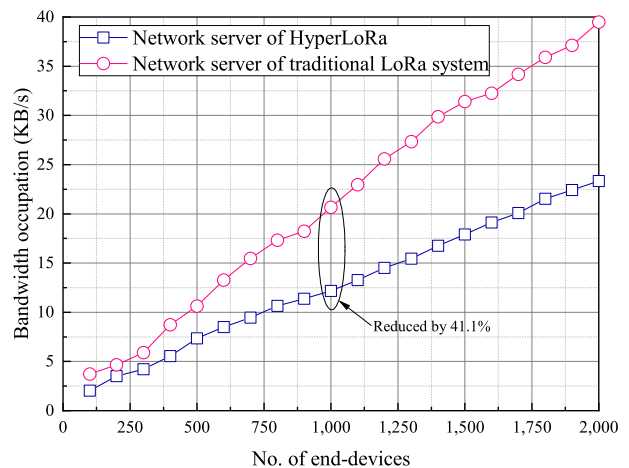}
	\caption{Illustration of bandwidth occupation of transmission links between LoRa gateways and network server in \textit{Experiment \ref{enum:3}}.}
	\label{bandwidth}
	\end{figure}

\begin{figure}[!h]
	\centering
	\includegraphics[width=0.45\textwidth]{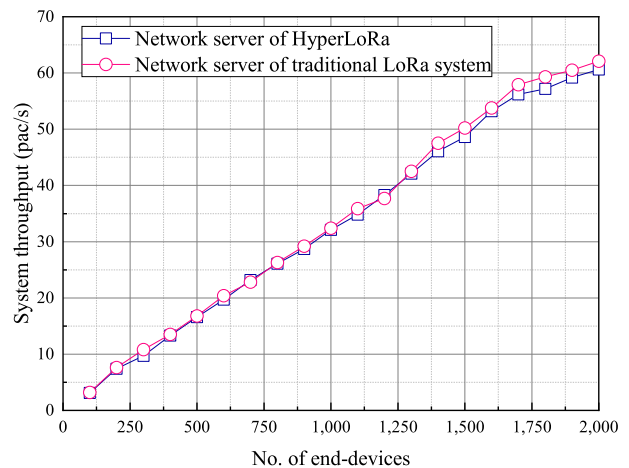}
	\caption{Illustration of system throughput with both valid and invalid application packages in \textit{Experiment \ref{enum:3}}.}
	\label{rps}
	\end{figure}

	Figure~\ref{cpumix} shows the CPU utilization of both LoRa gateways and network server in \textit{Experiment 3} and the comparison experiment. With the increase of end-devices, more CPU resources are occupied both in LoRa gateways and network server. Because LoRa gateways have processed all packages and filtered invalid packages, the CPU utilization of network server has reduced by $22\%$ compared with the network server of traditional LoRa system where network server needs to handle all packages itself. Not only CPU resources are spared, but also the bandwidth of transmission link between LoRa gateways and network server is saved, which is shown in Fig.~\ref{bandwidth}. This is because the invalid traffic has been blocked by LoRa gateways. The result shows that nearly $41.1\%$ save of bandwidth is gained when the number of end-devices is $1,000$, compared with traditional LoRa system. The gap keeps enlarging linearly as the increasing of end-devices. One of the benefits of the resource saving is that the spared resources can be used on more valid packages, avoiding them from being affected by the invalid or malicious traffic. Figure~\ref{rps} shows that the system throughput of the HyperLoRa is equivalent to the system throughput of traditional LoRa system. It means that HyperLoRa can reach the same performance as traditional LoRa system with less occupation of resources, owing to the efficient utilization of the resources in the LoRa gateways.
	
	\section{Conclusion}
	\label{sec:conclusion}
	This paper has proposed a new design of HyperLoRa system with edge computing, where a blockchain is deployed with two ledgers that store different types of LoRa data. Our proposed system can well utilize the limited resources at LoRa gateways to maintain one ledger. Two functions of network servers are also migrated to LoRa gateways to exploit the edge computing abilities. In HyperLoRa, some potential security risks can be mitigated or prevented by edge computing and the deployment of blockchain. Moreover, a prototype of HyperLoRa has been successfully implemented which demonstrates the feasibility of our proposed design. Through the experiments, we have shown the maximum performances that HyperLoRa can reach, e.g., handling almost $1,350$ join requests with \SI{5}{\second} timeout, and processing $48$ application packages per second with $1,000$ end-devices that upload data in \SI{15}{\second} interval. Furthermore, the central cloud can save CPU utilization and bandwidth occupation without any decrease on system throughput, compared with traditional LoRa system.
	
\bibliographystyle{IEEEtran}
\bibliography{ref}

\end{document}